\documentclass[aps,prc,superscriptaddress,showpacs]{revtex4}

\usepackage{graphicx}

\begin{document}

\title{ 
Isoscaling Studies of Fission - 
a Sensitive Probe into the Dynamics of Scission
      }

\author{M. Veselsky}
\email{fyzimarv@savba.sk}
\affiliation{Institute of Physics of the Slovak Academy of 
Sciences, Dubravska 9, Bratislava, Slovakia}
\affiliation{Cyclotron Institute, Texas A\&M University, College Station, 
TX 77843}

\author{G. A. Souliotis}

\affiliation{Cyclotron Institute, Texas A\&M University, College Station, 
TX 77843}

\author{M. Jandel}
\affiliation{Institute of Physics of the Slovak Academy of 
Sciences, Dubravska 9, Bratislava, Slovakia}

\date{\today}

\nopagebreak

\begin{abstract}
The fragment yield ratios were investigated in 
the fission of $^{238,233}$U targets induced by 14 MeV neutrons. 
The isoscaling behavior was typically observed for the isotopic chains 
of fragments ranging from the proton-rich to the most neutron-rich ones. 
The observed high sensitivity of neutron-rich heavy fragments to the target 
neutron content suggests fission as a source of neutron-rich 
heavy nuclei for present and future rare ion beam facilities, allowing 
studies of nuclear properties towards the neutron drip-line and investigations 
of the conditions for nucleosynthesis of heavy nuclei. 
The breakdowns of the isoscaling behavior around N=62 and N=80 manifest 
the effect of two shell closures on the dynamics of scission. 
The shell closure around N=64 can be explained by the deformed shell. 
The investigation of isoscaling  in the spontaneous fission of $^{248,244}$Cm 
further supports such conclusion. The Z-dependence of the isoscaling parameter 
exhibits a structure which can be possibly related to details of scission 
dynamics. The fission isoscaling studies can be a suitable tool for 
the investigation of possible new pathways to synthesize still heavier nuclei.

\end{abstract}



\maketitle

\section{Introduction}

Nuclear fission, first observed in 1938 \cite{Hahn,Meit}, has been investigated 
extensively for many decades \cite{Specht,Brosa,Hamilton}. The model 
description of scission ranged from the early statistical model 
of Fong \cite{Fong} to the advanced statistical model of Wilkins \cite{Wilkins} 
and, more recently, to the dynamical description 
of the random-neck rupture model of Brosa \cite{Brosa}. Despite the obtained 
level of understanding, there are still many open questions and 
challenges remaining ( for an overview see e.g. \cite{StAndrews}). 
Fission-like phenomena have 
been observed recently also in metallic clusters \cite{Naher}. 
Typically, nuclear fission 
produces a wide range of fragments with different atomic and mass 
numbers. Similar wide mass and charge distributions of reaction 
products can be observed in nuclear reactions leading to the production of 
hot nuclei which de-excite by emission of charged fragments. 
It has recently been observed  \cite{Tsang1} that for two similar reactions 
occurring at the same temperature, that differ only in the isospin asymmetry, 
the ratio R$_{21}$(N,Z) of the yields of a given fragment (N,Z) exhibits 
an exponential dependence on N and  Z of the form 
$R_{21}(N,Z) \propto  \exp(\alpha N + \beta Z)$ , 
this scaling behavior being termed isoscaling 
\cite{Tsang1}. Initially, the investigations of isoscaling focused on the 
yields of light fragments up to oxygen, originating from the massive hot 
systems produced at intermediate energies \cite{Tsang1}, or high energies 
\cite{IsoBotvina}. The isoscaling behavior was attributed to the difference 
of statistical de-excitation of two massive hot systems with different 
isospin asymmetry. In recent articles \cite{GSHRIso,GSHRIso2}, 
the investigation of isoscaling is reported using the heavy residue 
data from the reactions of 25 MeV/nucleon $^{86}$Kr projectiles 
with  $^{124}$Sn,$^{112}$Sn and $^{64}$Ni, $^{58}$Ni targets. 
The global isoscaling behavior is observed for heavy residues originating 
from the most damped collisions, while the dependence 
of the isoscaling parameters  $\alpha$ and  $\beta$ on deposited 
excitation energy can be used to extract information on the level 
of N/Z-equilibration between the projectile and target, ranging from 
no equilibration at the most peripheral collisions to full equilibration 
at the most damped collisions. The investigation of isoscaling 
on reconstructed hot quasi-projectiles with mass A=20-30 from the 
reactions of 30 and 50 MeV/nucleon $^{28}$Si 
with  $^{124,112}$Sn targets was reported in \cite{MVSiSnScal}. 
The observed dependence of the isoscaling parameter on excitation energy 
of the quasi-projectile was independent of the projectile energy and the 
resulting difference of inclusive isoscaling parameters at different 
projectile energies reflects different excitation energy distributions. 
Furthermore, due to the known level of N/Z-equilibration \cite{MVSiSnScal}, 
the temperature dependence of the isoscaling parameter 
was used to extract information on the onset of chemical separation. 
In general, the isoscaling studies 
appear to be a rather global method to investigate nuclear processes, 
manifesting the response of the system to the variation of the isospin 
asymmetry. The exponential scaling is typically a signal that 
some degree of equilibrium was reached. 

It is of interest to further explore the isoscaling 
across the wide range of nuclear processes, even the most 
traditional ones explored in most detail. In this study, we present 
the results of an isoscaling analysis 
of the fragment yield data from fission induced by fast neutrons. 
The evaluated independent fragment yields from the fission 
of $^{233,238}$U targets induced by 14 MeV neutrons \cite{ENDF349} 
were used. 



    \begin{figure}[h]                                        

   \includegraphics[width=0.66\textwidth, height=0.5\textheight ]{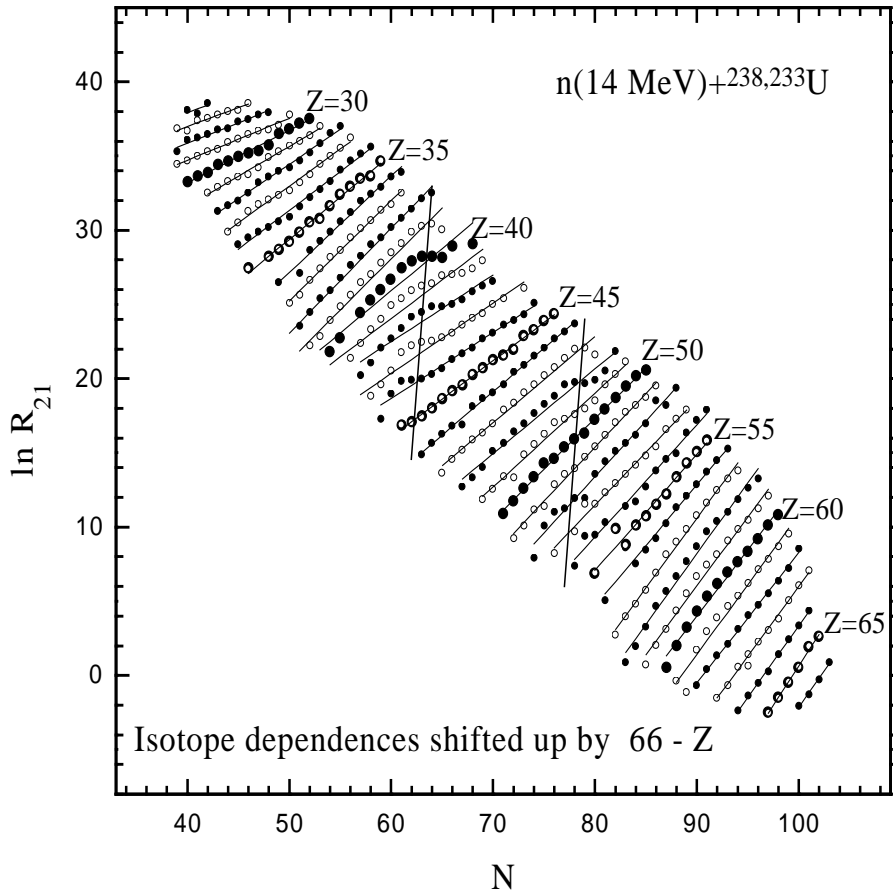}

    \caption{
Ratios of the fragment yields 
from the fission of $^{238,233}$U targets induced 
by 14 MeV neutrons \cite{ENDF349}. The data are shown as 
alternating solid and open circles. The labels apply to the larger 
symbols. The lines represent exponential fits. For clarity, the 
R$_{21}$ dependences are shifted from element to element by one unit. 
Nearly vertical lines mark major isoscaling breakdowns. 
           }
    \label{FScl}
    \end{figure}


\section{Isoscaling Analysis of Fission Fragment Yields}

In Fig. \ref{FScl}, we present the ratios of the fragment yields 
from the fission of $^{238,233}$U targets induced 
by 14 MeV neutrons, obtained using the recommended independent 
fission fragment yields from the evaluated nuclear data file 
ENDF-349 \cite{ENDF349}. The recommended yields are obtained 
as a result of the evaluation procedure and comprise both the fission 
yield measurements reported in the literature and the calculated Gaussian 
extrapolations of charge distributions for given masses. Despite 
limitations, such data possess a predictive power worth to explore, 
partially also due to the immensity and diversity of the existing 
experimental data. The study was restricted to recommended independent yields 
larger than 10$^{-5}$ \%.

In Fig. \ref{FScl}, one can observe, in the majority of the cases, a remarkable 
isoscaling behavior of isotopic chains with more than ten members. 
As mentioned above, some of the last data points are the results 
of extrapolation using the calculated Gaussian charge distributions. 
Nevertheless, since the evaluations have been done on a case-by-case 
basis, independently for each mass ( and not for each element ), 
the continuation of the trend they represent is possibly related 
to the physical behavior 
of the system without being excessively biased by the evaluation procedure. 
The available neutron multiplicity data from photofission 
of $^{238}$U at comparable excitation energy \cite{U238PhFiss} 
show that the fragments typically emit only two neutrons on average 
and thus the effect of emission is rather limited. Thus  
the observed behavior in Fig. \ref{FScl} may imply that the isoscaling 
behavior is mainly determined by the dynamical evolution of the system 
prior to scission.  

Of special interest for investigation are the mass regions 
corresponding to the most asymmetric mass splits. Practically all 
fragments below Zn exhibit weak dependence of R$_{21}$ on the neutron 
number, strongly contrasting the enhanced sensitivity of R$_{21}$  for 
their partners, the heaviest observed fission fragments. 
Such a weak dependence may imply a higher temperature of the 
scissioning light pre-fragment, since the isoscaling parameter is 
typically, within the grand-canonical approximation, considered inversely 
proportional to the temperature ( $\alpha$=$\Delta \mu_{n}$/T, 
$\Delta \mu_{n}$ being the difference of neutron chemical potentials 
of the two fragmenting systems ). 
It is, however not a priori obvious why such grand-canonical 
formula could be applied to fission. On the other hand, one can arrive at a 
formula of similar structure using the nucleon transport theory appropriate 
for fission. Typically the process of fission can be divided into 
two phases. The fissioning system must first overcome the saddle point 
( the peak of the fission barrier ), thus entering the irreversible path 
towards scission. In the second stage, the properties of the scission 
configuration 
are determined during the long descent from the saddle to the scission 
configuration. The process of fission ( as well as the closely related 
process of deep-inelastic nucleus-nucleus collision ) can be described 
in terms of collective motion using the transport theory 
\cite{Randrup,Feldmeier,HHofmann,Frobrich}. The dynamics 
of the collective degrees of freedom is typically described 
using the Langevin or Fokker-Planck equation. From a practical point of view, 
the isoscaling occurs when the two mass distributions for a given Z from 
two processes with different isospin are Gaussian distributions 
of the same width with different mean masses. 
Isotopic yield distributions can be, according to transport theory, 
considered as the solutions of the Fokker-Planck equation 
( Gaussian mass distributions ). A necessary condition for isoscaling, 
the assumption of equal width ( typically a sign of equal temperatures )  
leads in the context of the Fokker-Planck equation to equal values 
of the diffusion coefficients. Then the isotopic dependence of the 
yield ratio $R_{21}$ will assume the form 

\begin{equation}
R_{21}(N) \propto  \exp(\frac{V_{N2}-V_{N1}}{2D_{NN}} N ) 
\label{eqn0}
\end{equation}

where $V_{N1,2}$ are the drift coefficients characterizing the two processes 
with different isospin and $D_{NN}$ is the common diffusion coefficient. 
The transport theory typically complies with the fluctuation-dissipation 
theorem $V_{N}T = D_{NN}F_{N}$  which after expressing the driving force 
as $F_{N}=\Delta\mu_{nab}$ \cite{Feldmeier} leads to the following expression 
for the isoscaling coefficient 

\begin{equation}
\alpha=\Delta (\Delta\mu_{nab})/2T 
\label{eqn1}
\end{equation}

where $\Delta\mu_{nab}$ is the 
difference of neutron chemical potentials of interacting nuclei $a,b$ 
determining the drift velocity. Thus apparently the isoscaling behavior is 
consistent with statistical transport theory. Within the statistical transport 
theory, the two interacting nuclei ( fission fragments ) can both be treated 
statistically using a grand-canonical approach \cite{Feldmeier}, with two 
different temperatures, with the stochastic force depending in the 
general case on both temperatures. The temperature $T$ thus possibly can be 
related to the Fermi gas temperature of the fragments in the given 
isotopic chain. In the photofission of $^{238}$U, 
the observed neutron multiplicities for the lightest and heaviest fragments 
observed ( about two neutrons on average \cite{U238PhFiss} for A$\approx$90 
and A$\approx$150 ) 
correspond to different fragment temperatures, higher temperature 
for the lightest fragments and lower for the heaviest fragments. 
In Fig. \ref{FScl} the light fragments typically exhibit weak 
logarithmic slopes while their heavy partners exhibit much larger logarithmic 
slopes. The equation (\ref{eqn1}), based on the transport theory, 
suggests that the logarithmic slopes ( isoscaling parameters ) can indeed 
be related to temperature and excitation energy, thus possibly establishing 
an analogy with \cite{U238PhFiss}. A quantitative test of 
the equation (\ref{eqn1}) would be helpful in order to establish such analogy. 

The high sensitivity of neutron-rich heavy fragments to the neutron content 
of the fissioning system, combined  with observation from \cite{U238PhFiss} 
that the multiplicity of neutrons emitted from heavier fragment is less 
dependent on photon energy, is of considerable interest for the production 
of neutron-rich nuclei, since it suggests that neutron-rich 
heavy fragments can be created in the fission of neutron-rich uranium isotopes, 
thus leading to the production of neutron-rich nuclei which 
are of crucial importance for future rare isotope beam facilities, e.g. 
\cite{RIAANL,RIAMSU,EurIsol}. From a nuclear reaction standpoint, a 
promising candidate for the production 
of very neutron-rich fissile nuclei would be the peripheral 
collision of heavy fissile nuclei at energies between the Coulomb barrier 
and the Fermi energy \cite{GSEmis} leading to the moderately excited 
heavy uranium-like nuclei which would further fission into neutron-rich 
fragments. Some encouraging results have been obtained already using 
a $^{238}$U beam at 20 MeV/nucleon \cite{GSfission,GSStAndrews}. 
Of special interest for future experiments is the reaction of 
$^{238}$U beam with targets exhibiting the effect of neutron skin, 
such as $^{64}$Ni and $^{124}$Sn \cite{GS_PLB,GS_PRL} or $^{208}$Pb and 
$^{238}$U. 

Apart from the typically regular isoscaling behavior across the 
periodic table, the isoscaling behavior is violated in some regions, 
most prominently around N=62, around N=80 and possibly around N=86-88. 
The violation of the isoscaling in the regions N=80 
can be possibly related to the neutron shell closure around N=82. 
The isoscaling behavior around another strong neutron shell closure 
around N=50 does not seem to exhibit irregularities exceeding the 
statistical ones. The breakdown of isoscaling around N=62 
can not be directly related to the effect of any spherical shell closure 
in the final fission fragments. On the contrary, in this region the nuclei 
are supposed to be typically deformed. However, according 
to the fission model of Wilkins et al. \cite{Wilkins}, it may be related 
to the effect of the deformed neutron shell at N=64 with quadrupole 
deformation $\beta_{quad}$=0.6 ( point C in Fig. 1 of \cite{Wilkins} ) 
which plays a crucial role in the dynamics of 
scission. In a similar manner the violation of isoscaling in the region 
N=80 and possibly N=86-88 can be related to points G ( N=82 
and $\beta_{quad}$=0 ) and H ( N=88 and $\beta_{quad}$=0.65 ). 
The smooth dependence for Z=50 can be explained by the influence 
of spherical proton shell Z=50, again in accord with the conclusions 
of \cite{Wilkins}. Thus the observed isoscaling behavior appears 
to be generally consistent with the model calculations of Wilkins 
et al. concerning the scission configuration 
which allows to assume that it may reflect the real properties 
of the scission configuration. On the other hand, the probability 
of a given fission channel can be considered as a function of the 
scission configuration as a whole. Within the grand-canonical picture 
of isoscaling which describes de-excitation phenomena \cite{Tsang1}, 
the effect of binding energy ( and thus of the ground state shell correction ) 
of the final product is canceled out and only the properties of the whole 
excited system are reflected. In analogy, since the scission configuration 
is essentially binary, one can assume that the irregularities exhibited 
for final fragments around some neutron number may be caused by the 
shell structure of the other fragment. Then the observed structure around 
N=62 can be possibly related to the effect of N=82 shell closure 
in the other fragment. However, in the same sense the observed structure 
around N=80 should not be attributed to the shell closure around N=82 
but to the presence of shell closure around approximately N=64 of the 
other fragment. Thus, again the observed behavior points to the presence 
of the dominant shell closures in the scission configuration, one around N=82 
and another one around N=62-64. The former one can be either 
the ground state spherical shell closure or the deformed shell 
closure as calculated by Wilkins et al. \cite{Wilkins}. The latter 
shell closure can be only deformed ( as calculated by Wilkins 
et al. \cite{Wilkins} ). In any case the mass of the fissioning 
systems $^{234,239}$U causes that the two observed effects are exhibited 
by complementary fission fragments.


    \begin{figure}[h]                                        

   \includegraphics[width=0.66\textwidth, height=0.5\textheight ]{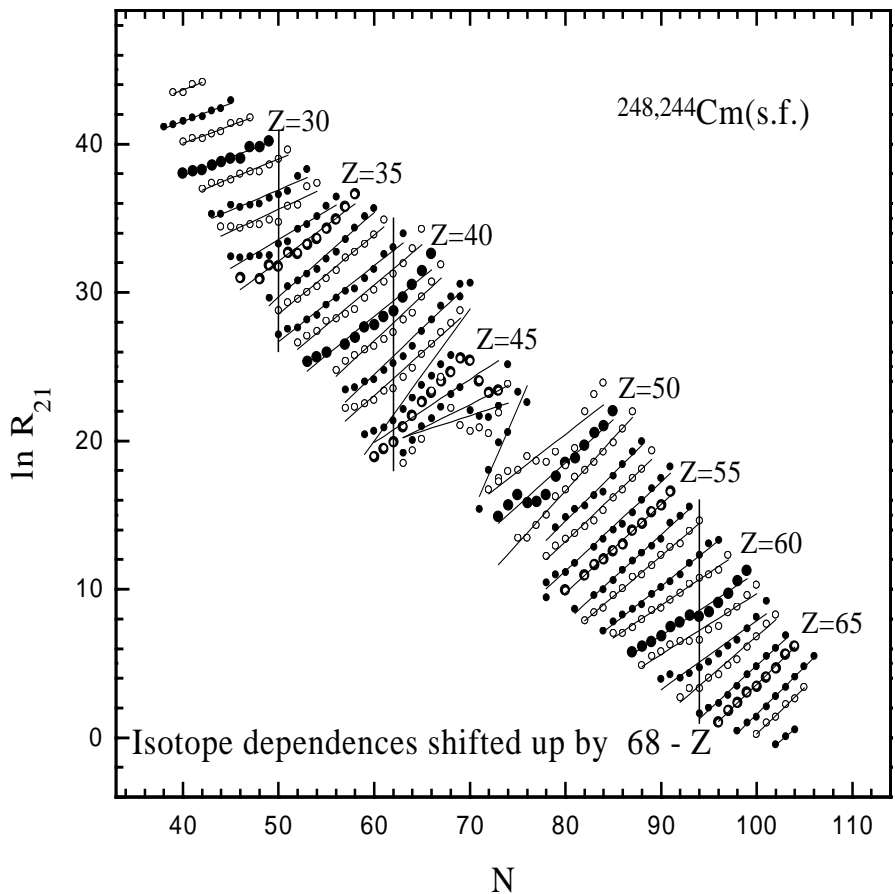}

    \caption{
Ratios of the fragment yields 
from the spontaneous fission of $^{248,244}$Cm \cite{ENDF349}. 
The data are shown as 
alternating solid and open circles. The labels apply to the larger 
symbols. The lines represent exponential fits. For clarity, the 
R$_{21}$ dependences are shifted from element to element by one unit. 
Vertical lines mark major isoscaling breakdowns. 
           }
    \label{CmFIscl}
    \end{figure}


In order to separate the two effects, a logical step is to continue the 
analysis by investigating a fissioning system 
of different mass, for practical reasons heavier. 
The systematic fission data for nuclei heavier than uranium 
are rather scarce, one nevertheless can find a good candidate in 
evaluated fragment yield data from spontaneous fission of heavier 
nuclei such as $^{248,244}$Cm \cite{ENDF349}. The spontaneous fission is a much 
colder process where additional phenomena such as barrier penetration 
probability and potential energy surface can play a crucial role, 
it is nevertheless interesting to explore it in the context of isoscaling. 
The isoscaling plot from the spontaneous fission of $^{248,244}$Cm 
is shown in Fig. \ref{CmFIscl}. 
In the spontaneous fission of $^{248,244}$Cm, the overall isoscaling behavior 
can be observed from the most populated region of mass distributions towards 
asymmetric mass splits. 
The breakdowns featuring rather the slope change than the structures 
observed in the previous case are observed around N=50 and N=62. 
A significant effect is observed around N=68-70 ( with complementary fragment 
around N=82 ), which nevertheless can be a sign of transition 
into symmetric region around N=74-76 where the behavior 
is rather irregular. Further the isoscaling behavior is quite regular 
up to the heaviest fragments ( with the exception around N=94 where 
a hint of isoscaling breakdown can be observed ). It is remarkable that the 
breakdown of isoscaling ( slope change ) is again observed around N=62 
( along with N=50 ) which possibly points to the role of shell structure, 
which may influence the potential energy surface and thus the probability 
of a given fission channel. While around N=50 the shell structure can 
be identified with spherical fragment, for the region around N=62 
the influence of the deformed shell is the most plausible. 
The observed isoscaling behavior in the spontaneous fission of $^{248,244}$Cm, 
despite the more complex dynamics of the cold process, again appears 
to point to the role of the deformed shell closures ( in particluar 
around N=64 ) in the fission of actinides.


    \begin{figure}[h]                                        

   \includegraphics[width=0.44\textwidth, height=0.44\textheight ]{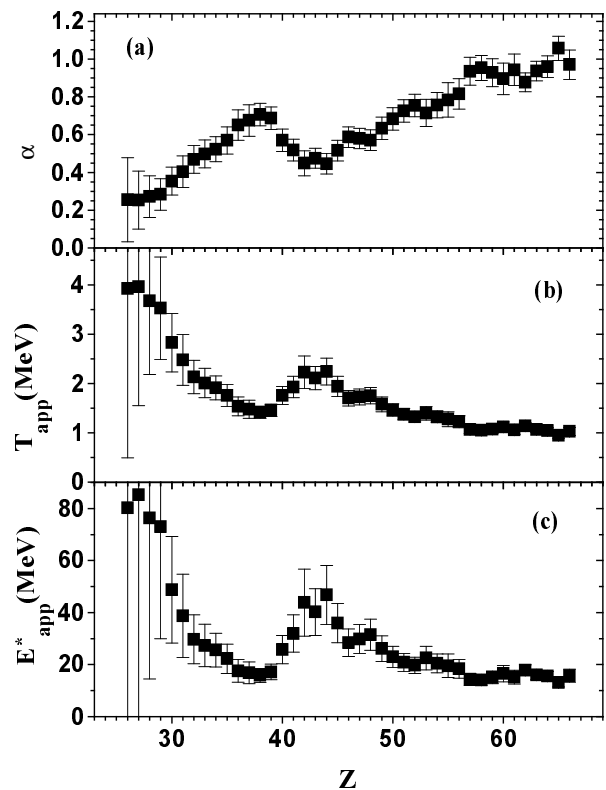}

    \caption{
The systematics of (a) isoscaling parameter $\alpha$ in the fission 
of $^{238,233}$U targets induced by 14 MeV neutrons 
(b) estimates of apparent fragment temperatures obtained using 
Eqn. (\ref{eqn1}) ( for details see text ) and (c) corresponding 
fragment excitation energies as a function of the atomic number Z. 
           }
    \label{avsZ}
    \end{figure}


The above isoscaling analysis was carried out using the recommended 
fission fragment yields obtained as a compilation 
of the existing fission fragment yield data complemented with 
model based extrapolations. The uncertainty of the extrapolated 
data depends on the detailed extrapolation procedure and its magnitude 
is not expressed quantititively. It is of interest to establish whether the 
observed behavior could be influenced by the uncertainty of the extrapolation. 
Since both the overall isoscaling behavior and the irregularities 
in particular regions around the neutron numbers N=62 and N=80 
are observed across many isotopic chains with values of yields ranging 
across several orders of magnitude, one can conclude that the observed 
behavior corresponds to the details of the fission process rather than 
to the uncertainty of the data. 

For the fission of $^{238,233}$U targets induced by 14 MeV neutrons, 
we present in Fig. \ref{avsZ}a the systematics of the isoscaling parameters 
$\alpha$, obtained by exponential fits of the isotope chains, 
as a function of the atomic number Z. One line was fitted for each Z 
( see Fig. \ref{FScl} ). As already discussed, 
the isoscaling parameter $\alpha$ decreases below Z=32, while one can 
observe an increase above Z=58, thus possibly signaling a relatively 
hot light fragment and a colder heavy fragment in the asymmetric fission 
channels. In the central part of the fragment mass distribution, one can 
observe an increase up to the maximum around Z=38, then 
an abrupt discontinuity around Z=40 and a linear increase above. 
According to formula (\ref{eqn1}), the decrease of the isoscaling parameter 
$\alpha$ around Z=42 signals rather hot ( and possibly initially 
deformed ) fragments which is a reasonable result when considering that the 
second fragment for such a mass split is in the vicinity of Z=50 proton shell 
and it can be expected to be spherical and possibly colder, as 
one can conclude from the increase of values of $\alpha$ around Z=50. 
This is again consistent with the model calculations of Wilkins et al. 
\cite{Wilkins} which obtained a decrease of distance of centers 
for corresponding mass splits ( see Fig. 27 of \cite{Wilkins} ). 
Also the recent results from ternary 
fission suggest that typically the low energy light charged 
particle in the spontaneous fission of $^{252}$Cf is correlated 
with a $^{132}$Sn fragment and a strongly deformed light complementary  
fragment at scission \cite{Jandel1,Jandel2}, that may de-excite 
by secondary emission of the low energy ternary particle. 
One can also attempt to relate the discontinuous Z-dependence of the 
isoscaling parameter $\alpha$ to the sawtooth dependence of the neutron 
multiplicity on fragment mass \cite{U238PhFiss,Terrell,Bowman}, 
which motivated the model of random neck-rupture \cite{Brosa}. However, 
in the partially complete data from the photofission of $^{238}$U 
\cite{U238PhFiss}, the minimum  
in the neutron multiplicity, signaling the position of the discontinuity, 
appears at the mass approximately A=130 rather than at masses A=100-108, 
corresponding approximately to Z=42. The way for detailed comparison 
of the isoscaling parameters $\alpha$ with the neutron multiplicity, 
determined in \cite{U238PhFiss}, is not trivial, 
since there the measured  neutron multiplicity for a given fragment 
pair was divided between the fragments according to the model calculation. 
In any case, the observed discrepancy deserves further investigations. 

As a further step in the isoscaling analysis, we test the potential 
of the Eqn. (\ref{eqn1}), derived from transport theory, to provide 
quantitative information on the properties of the scission configuration. 
In Fig. \ref{avsZ}b,c we present an estimate of the apparent fragment 
temperatures obtained using Eqn. (\ref{eqn1}) and of the corresponding 
excitation energies. The quantity $\Delta(\Delta\mu_{nab})$ was approximated 
using the difference of neutron separation energies of two composite systems. 
Such approximation can be based on the assumptions that 
$\Delta(\Delta\mu_{nab})$, 
reflecting the difference of neutron transfer driving forces for two 
fissioning systems, will track with the difference of neutron chemical 
potential of the two composite systems. Thus we assume that the average driving 
force for a given fissile system is a property of the whole system and 
that its relative change between two fissiong systems is equal to 
the difference of chemical potentials, which can be roughly approximated 
by their neutron separation energies. However rough such approximation 
is, it provides an interesting quantitative test of Eqn. (\ref{eqn1}) 
and thus also of the underlying theory. As one can see in Fig. \ref{avsZ}b,c,  
the extracted values of the apparent temperature and excitation  energy indeed 
lead in some mass regions to reasonable agreement with the experimental neutron 
multiplicities of photofission of $^{238}U$ \cite{U238PhFiss} at comparable 
excitation energies. Specifically for the most asymmetric splits 
( A$\approx$90 vs 150, corresponding roughly to Z$\approx$36 vs 56 
in the present work ) observed in \cite{U238PhFiss} approximately two neutrons 
are emitted from each fragment, thus suggesting the fragment excitation 
energies of about 15-20 MeV. For more central mass splits some 
similarity of trends can be observed, especially the excitation energy 
increase with increasing mass of the light fragment. The observed rapid 
increase of estimated temperatures and excitation energies for the lightest 
fragments in Fig. \ref{avsZ}b,c may be caused by statistical uncertainties. 
While the above estimate is indeed rough, it nevertheless 
suggests a possibility to study details of the fission dynamics, e.g. 
to map the transfer driving force using the fragment temperatures measured 
independently. The specific observation that the estimated fragment 
temperatures are different for complementary light and heavy fragments can be 
understood naturally within the transport theory, where the 
number of transfers in both directions is typically comparable, 
thus leading to comparable excitation energies, which in the case 
of very asymmetric mass splits lead to a hotter light fragment 
and colder heavy fragment. Such a situation is routinely observed 
in peripheral mass-asymmetric nucleus-nucleus collisions ( see for example 
\cite{MVSiSn} ), where the light quasiprojectile can get very hot while 
the quasitarget remains relatively cold.  

In general, the isoscaling analysis of the fission 
data appears to be a sensitive tool to investigate the fission dynamics. 
While it can not substitute the traditional experimental investigations 
of fission, its simplicity can prove useful in situations where such 
studies are not possible. For instance, as suggested in \cite{MVHIPh02}, 
the knowledge of the fission dynamics of superheavy nuclei is essential 
for understanding of the possibility to synthesize still heavier 
nuclei. In this aspect, the isoscaling analysis of reactions with 
two targets, possibly isotopes of the same element, can allow 
investigations of the fission dynamics of very heavy nuclei, using either 
a designated experimental setup or radiochemical methods. From a theoretical 
point of view, the capability to reproduce the isoscaling behavior can be 
a simple but, nevertheless, crucial test of any fission model. Also, as shown 
in the above analysis, the isoscaling properties can provide essential 
information on the possibility to produce heavy neutron-rich 
nuclei which are essential to understand the nuclear 
properties in general and, in addition, the astrophysical processes 
leading to the nucleosynthesis of the heaviest elements.


\section{Summary and Conclusions}

In summary, the fragment yield ratios were investigated in 
the fission of $^{238,233}$U targets induced by 14 MeV neutrons. 
The isoscaling behavior was typically observed for isotopic chains ranging 
from the most proton-rich to most neutron-rich ones. The high sensitivity 
of the neutron-rich heavy fragments to the target 
neutron content suggests the viability of fission ( possibly following 
a peripheral collision with another n-rich nucleus ) as a source of very 
neutron-rich 
heavy nuclei for future rare ion beam facilities, thus allowing 
studies of the nuclear properties of such nuclei and exploration 
of the conditions for the nucleosynthesis of heavy nuclei.  The observed 
breakdowns of the isoscaling behavior around N=62 and N=80 indicate 
the effect of two major shell closures on the dynamics of scission, 
one of them being the deformed shell closure around N=64. The isoscaling 
analysis of the spontaneous fission of $^{248,244}$Cm further supports such 
conclusion. The values of the isoscaling parameter appear to exhibit 
a structure which can be possibly related to details of scission 
dynamics e.g. for the charge split Z=42 vs Z=50.  
The isoscaling studies present a suitable tool for investigation of the 
fission dynamics of the heaviest nuclei, which can provide essential 
information about possible pathways to the synthesis of still 
heavier nuclei. 

\section*{Ackowledgements}

\par

This work was supported through grant of Slovak Scientific Grant Agency 
VEGA-2/1132/21. The research at the Cyclotron Institute of Texas A\&M 
University is supported in part by the Robert A. 
Welch Foundation through grant No. A-1266, and the Department of Energy
through grant No. DE-FG03-93ER40773. 


\bibliography{fisiso4.bib}



\end{document}